\documentclass[aps,prl,twocolumn,showpacs,preprintnumbers,
               amsmath,amssymb]{revtex4}

\newcommand{\IM}{\mbox{\rm Im}}
\newcommand{\eqn}[1]{(\ref{#1})}
\newcommand{\mev}{\mbox{\rm MeV}}
\newcommand{\gev}{\mbox{\rm GeV}}

\begin{document}

\preprint{HD-THEP-02-01}

\title{Flavour-symmetry breaking of the quark condensate and chiral
 corrections to the Gell-Mann-Oakes-Renner relation}

\author{Matthias Jamin}
\email{jamin@uni-hd.de}
\altaffiliation{Heisenberg fellow.}
\affiliation{Institut f\"ur Theoretische Physik, Universit\"at Heidelberg,
Philosophenweg 16, D-69120 Heidelberg, Germany}

\date{February 1, 2002}

\begin{abstract}
The relation between the chiral quark condensate in QCD sum rules and chiral
perturbation theory is clarified with the help of a low-energy theorem for
the scalar and pseudoscalar correlation functions. It is found that the quark
condensate should be identified with the non-normal-ordered vacuum expectation
value of quark-antiquark fields. Utilising results on flavour SU(3) breaking
of the quark condensate from QCD sum rules, the low-energy constant $H_2^r$ in
the chiral Lagrangian, as well as next-to-leading order corrections to the
Gell-Mann-Oakes-Renner relation are estimated.
\end{abstract}


\maketitle

%%%%%%%%%%%%%%%%%%%%%%%%%%%%%%%%%%%%%%%%%%%%%%%%%%%%%%%%%%%%%%%%%%%%%%%%%
% The main part of the paper
%%%%%%%%%%%%%%%%%%%%%%%%%%%%%%%%%%%%%%%%%%%%%%%%%%%%%%%%%%%%%%%%%%%%%%%%%

\section{Introduction}

During the last two decades, QCD sum rules \cite{svz:79} and chiral perturbation
theory ($\chi$PT) \cite{gl:84} have become indispensable tools in the
phenomenology of particle physics. Both methods address the problem of dealing
with QCD in the non-perturbative low-energy regime. Whereas in QCD sum rules
the basic degrees of freedom are quarks and gluons and the non-perturbative
region is approached from higher energies by including vacuum averages of
composite operators, the condensates, $\chi$PT constitutes a low-energy
expansion with the lowest lying hadrons as the fundamental fields. Nevertheless,
a sound matching of QCD and low-energy effective theories still is an open
problem of QCD phenomenology \cite{ppr:98}.

$\chi$PT is constructed such as to reflect the symmetry properties of QCD
under chiral transformations. The QCD pattern of explicit chiral symmetry
breaking is implemented by adding quark mass terms in the $\chi$PT Lagrangian,
which introduces an additional free parameter $B_0$ at the leading order,
besides the pion decay constant $f_\pi$. Since the $\chi$PT mass term contains
the scalar quark density, it turns out that the parameter $B_0$ is related to
the quark condensate.

The quark condensate in full QCD is not directly related to a physical
quantity. Like the QCD coupling and the quark masses, its value depends on
the renormalisation scale and on the renormalisation scheme employed in
the calculation.\footnote{Contrary to the gluon condensate, the quark
condensate, being an order parameter of chiral symmetry breaking, does not
suffer from a renormalon ambiguity \cite{ben:99}.} On the other hand, in
$\chi$PT the information on the short-distance properties of the theory is
hidden in the low-energy constants, and thus the question arises, how the
quark condensates of QCD and $\chi$PT are related. The key to the answer lies
in a low-energy theorem for the scalar and pseudoscalar correlation functions,
which will be discussed in the next section.

In $\chi$PT, the dependence of the quark condensate on the renormalisation is
reflected in an unphysical low-energy constant $H_2^r$, which appears at the
next-to-leading order and is required for the renormalisation of the chiral
expansion. $H_2^r$ thus comprises information on short-distance physics, which
cannot be deduced from $\chi$PT alone. In section three, the constant $H_2^r$
will be estimated from flavour SU(3) breaking of the quark condensate. This
flavour breaking can for example be extracted from QCD sum rules for the
heavy-light $B$- and $B_s$-meson systems.

Besides the flavour-symmetry breaking of the quark condensate, $H_2^r$ also
governs the next-to-leading chiral correction to the well known
Gell-Mann-Oakes-Renner (GMOR) relation \cite{gmor:68}. The GMOR relation
constitutes the main source for numerical values of the quark condensate,
assuming the light quark masses to be known. Making use of the estimate of
$H_2^r$, in section four, also estimates for these chiral corrections and the
quark condensate will be presented.

\section{The scalar correlator}

The central object which is investigated in the seminal version of QCD sum
rules \cite{svz:79} is the two-point function $\Psi(p^2)$ of two hadronic
currents
\begin{equation}
\label{psip2}
\Psi(p^2) \equiv i \int \! dx \, e^{ipx} \,
\langle\Omega| \, T\{\,j(x)\,j(0)^\dagger\}|\Omega\rangle\,,
\end{equation}
where $\Omega$ denotes the physical vacuum and in our case $j(x)$ will be the
divergence of the flavour-changing vector (axialvector) current,
\begin{equation}
\label{j}
j(x) = \partial^\mu (\bar Q\gamma_\mu (\gamma_5)q)(x) =
i\,(M\mp m)(\bar Q(\gamma_5)q)(x) \,.
\end{equation}
$M$ and $m$ represent the masses of the quarks $Q$ and $q$ respectively, and
the upper (lower) sign always corresponds to the scalar (pseudoscalar) case.
Up to a factor $(M\mp m)^2$, $\Psi(p^2)$ is just given by the scalar
(pseudoscalar) two-point function. Furthermore, $\Psi(p^2)$ satisfies a
dispersion relation with two subtractions,
\begin{equation}
\label{disrel}
\Psi(p^2) = \Psi(0) + p^2\,\Psi'(0) + p^4\!\int\limits_0^\infty
\frac{\rho(s)}{s^2(s-p^2-i0)}\,ds \,,
\end{equation}
where $\rho(s)\equiv\IM\,\Psi(s+i0)/\pi$ is the spectral function corresponding
to $\Psi(s)$.

A low-energy theorem relates the first subtraction constant $\Psi(0)$ to the
product of quark masses and quark condensates:
\begin{equation}
\label{lot}
\Psi(0) = -\,(M\mp m)(\,\langle\Omega|\bar QQ|\Omega\rangle \mp
\langle\Omega|\bar qq|\Omega\rangle\,) \,.
\end{equation}
The relation \eqn{lot} can be derived by taking the divergence of the vector
(axialvector) correlator, employing current commutation relations in addition
\cite{bro:75}. Up to the next-to-leading order in the strong coupling constant,
it was also verified in QCD perturbation theory \cite{bro:81}.

From the time-ordered product in eq.~\eqn{psip2} and Wick's theorem, one
naturally obtains normal-ordered condensates. However, in the calculation of
$\Psi(0)$ also quartic mass corrections as well as higher-dimensional operators
arise. On dimensional grounds, the operator corrections with dimension greater
than four are accompanied by inverse powers of the quark masses thus being
singular in the chiral limit. If the normal-ordered quark condensates are
rewritten in terms of the {\em non-normal-ordered} minimally-subtracted
condensates, the quartic mass as well as higher-dimensional operator
corrections are cancelled, such as to yield the simple functional form of
eq.~\eqn{lot} \cite{bro:81,sc:88,jm:93,cdps:95}.

The fact that the quark condensates in eq.~\eqn{lot} should be considered as
{\em non-normal-ordered} implies that the subtraction constant $\Psi(0)$
depends on the renormalisation scale and scheme. Whereas the product of quark
mass times normal-ordered condensate is a renormalisation invariant quantity
\cite{mqq}, this is no longer true for the non-normal-ordered condensate. In
the latter case, the renormalisation invariant involves additional quartic
mass terms \cite{sc:88}. Numerically, the quartic mass terms only have some
relevance for the strange quark. For the lighter up and down quarks they are
completely negligible.

By considering QCD sum rules in which the subtraction constant $\Psi(0)$
remains \cite{psi0}, it would in principle be possible to determine the quark
condensate or flavour-breaking ratios of condensates. Here, care has to
be taken that the quartic mass corrections on the left-hand side of eq.
\eqn{disrel}, which cancel the renormalisation dependence of $\Psi(0)$ are
included properly. Nevertheless, even though the various quantities related
to $\Psi(0)$ generally were found to be reasonable, since the sum rules in
question are subject to very large higher-order perturbative corrections
\cite{jop:01}, these results should be interpreted with caution.

\section{The quark condensate in $\chi$PT}

The status of the quark condensate in $\chi$PT can be clarified by calculating
the scalar (pseudoscalar) correlator and the corresponding low-energy theorem
also in this framework \cite{gl:84}. It is then found that the functional
form of $\Psi(0)$ exactly resembles eq.~\eqn{lot}, and thus the quark
condensate in $\chi$PT should be identified with the {\em non-normal-ordered}
vacuum average of quark-antiquark fields.

As has been already discussed in the introduction, in $\chi$PT the quark
condensate, as well as flavour-breaking ratios, depend on the unphysical
low-energy constant $H_2^r$. An example of this dependence is displayed by
the ratio
\begin{equation}
\label{ssoqq}
\frac{\langle\bar ss\rangle}{\langle\bar qq\rangle} = 1 + 3\mu_\pi -
2\mu_K - \mu_\eta + \frac{8\Delta_{K\pi}}{f_\pi^2}\,(2L_8^r + H_2^r) \,,
\end{equation}
where $\langle\bar qq\rangle\equiv[\langle\bar uu\rangle+\langle\bar dd\rangle]
/2$ represents the isospin average of the up- and down-quark condensates, and
in the following, the vacuum state $\Omega$ is omitted. Furthermore, the
$\mu_P$ are chiral logarithms which take the form
\begin{equation}
\label{muP}
\mu_P = \frac{M_P^2}{32\pi^2 f_\pi^2}\,\ln\frac{M_P^2}{\nu_\chi^2} \,,
\end{equation}
with $\nu_\chi$ being the chiral renormalisation scale, $\Delta_{K\pi}=M_K^2-
M_\pi^2$, and $L_8^r$ is a physical low-energy constant in the next-to-leading
order chiral Lagrangian \cite{gl:84}.

The relation \eqn{ssoqq} offers the possibility to determine the low-energy
constant $H_2^r$, since the ratio $\langle\bar ss\rangle/\langle\bar qq\rangle$
can be obtained independently, e.g.~in the framework of QCD sum rules. A rather
reliable place to determine $\langle\bar ss\rangle/\langle\bar qq\rangle$ are
sum rules for the leptonic decay constants of the heavy $B$- and $B_s$-mesons
\cite{jl:01} (and references therein), since the deviation of the flavour SU(3)
breaking ratio $f_{B_s}/f_B$ from one is very sensitive to the quark condensate
ratio $\langle\bar ss\rangle/\langle\bar qq\rangle$.

The ratio of the heavy meson decay constants $f_{B_s}/f_B$ is, unfortunately,
not yet known experimentally. Thus we have to resort to a different framework
to find results on this ratio, independent of QCD sum rules. Such results
have been obtained in the framework of lattice QCD and a recent average yielded
$f_{B_s}/f_B=1.16\pm 0.04$ \cite{ber:00}.\footnote{More recent results can be
found in refs.~\cite{fblat}.} Employing this value in the QCD sum rules, the
corresponding flavour breaking ratio of the quark condensates turns out to be
\begin{equation}
\label{ssoqqSR}
\langle\bar ss\rangle / \langle\bar qq\rangle = 0.8 \pm 0.3 \,.
\end{equation}
Within the uncertainties, this result is in agreement to previous
determinations \cite{ssoqq}. Nevertheless, older results have not been included
in eq.~\eqn{ssoqqSR}, since in these investigations normal-ordered condensates
have been used and the corresponding quark mass corrections have not been
included. A proper treatment of the determination of $\langle\bar ss\rangle/
\langle\bar qq\rangle$ from QCD sum rules for other channels will be left for
future work.

In addition to the ratio $\langle\bar ss\rangle/\langle\bar qq\rangle$, for the
determination of $H_2^r$ also the low-energy constant $L_8^r$ is required.
This constant can be determined from flavour-symmetry breaking of meson masses
and decay constants, which at next-to-leading order in $\chi$PT satisfies the
relation \cite{gl:84}
\begin{equation}
\label{L8r}
\frac{1}{(R+1)}\,\frac{M_K^2}{M_\pi^2} + \frac{f_K}{f_\pi} = \frac{3}{2} +
\frac{3}{4}\,\mu_\pi - \frac{1}{2}\,\mu_K - \frac{1}{4}\,\mu_\eta +
\frac{8\Delta_{K\pi}}{f_\pi^2}\,L_8^r \,.
\end{equation}
Here $R$ represents the quark-mass ratio $R\equiv m_s/\hat m=24.4\pm 1.5$
\cite{leu:96}, and the value $f_K/f_\pi=1.22\pm 0.01$ \cite{pdg:00} will be
used. Employing the relation \eqn{L8r}, the constant $L_8^r$ is found to be
\begin{equation}
\label{L8rv}
L_8^r(\nu_\chi\!=\!M_\rho) = (0.88\pm 0.24)\cdot 10^{-3} \,,
\end{equation}
where the pion decay constant has been varied in the range $f_\pi=92.4\pm 10.0
\;\mev$. Although the experimental uncertainty for $f_\pi$ is much smaller
\cite{pdg:00}, the chosen error reflects the difference between the physical
value and the leading order decay constant $f_0=82\;\mev$, such that
uncertainties from higher orders in the chiral expansion should be largely
taken into account. The result \eqn{L8rv} is in agreement to the original
value obtained by Gasser and Leutwyler \cite{gl:84} at the scale
$\nu_\chi\!=\!M_\eta$.

The finding of eq.~\eqn{L8rv} for $L_8^r$ can also be compared with the very
recent value $L_8^r(\nu_\chi\!=\!M_\rho)=(0.62\pm 0.20)\cdot 10^{-3}$
\cite{abt:01}, which has been obtained in $\chi$PT at ${\cal O}(p^6)$. Within
the uncertainties both results are compatible, but it is seen that including
the next-next-to-leading corrections in the chiral expansion effectively lowers
the constant $L_8^r$. Since the presented analysis stays at the next-to-leading
order, for consistency also the corresponding value \eqn{L8rv} for $L_8^r$ has
been utilised in the following.

Inserting the results of eqs.~\eqn{ssoqqSR} and \eqn{L8rv} into \eqn{ssoqq},
it is possible to deduce an estimate on the unphysical low-energy constant
$H_2^r$:
\begin{equation}
\label{H2r}
H_2^r(\nu_\chi\!=\!M_\rho) = (-\,3.4\pm 1.5)\cdot 10^{-3} \,.
\end{equation}
The negative value for $H_2^r(\nu_\chi\!=\!M_\rho)$, found in eq.~\eqn{H2r}, is
in contrast to estimates of the same quantity in the framework of $\chi$PT with
explicit inclusion of resonance fields \cite{egpr:89}. Assuming saturation
of the low-energy constants in the chiral Lagrangian by meson resonances, 
reasonable values for the physical constants $L_i^r$ are obtained, whereas in
the case of $H_2^r$, saturation with scalar resonances yields the relation
$H_2^r=2L_8^r$, which is clearly violated by the result \eqn{H2r}.\footnote{In
fact, employing the relation $H_2^r=2L_8^r$, lead the authors of \cite{abt:01}
to deduce $\langle \bar ss\rangle/\langle\bar qq\rangle = 1.69$.} Since $H_2^r$
and $2L_8^r$ have the same dependence on the chiral scale $\nu_\chi$, it is
even impossible to find a particular scale at which the relation $H_2^r=2L_8^r$
would hold.

\section{Chiral corrections to the Gell-Mann-Oakes-Renner relation}

The unphysical low-energy constant $H_2^r$ also appears in the next-to-leading
order chiral corrections to the GMOR relations \cite{gmor:68,gl:84},
\begin{eqnarray}
\label{gmor1}
4\hat m\langle \bar qq\rangle &\!\!=\!\!& -\,2f_\pi^2 M_\pi^2\,(\,1-\delta_\pi
\,) \,, \\[1mm]
\label{gmor2}
(m_s+\hat m)[\,\langle\bar ss\rangle + \langle\bar qq\rangle\,] &\!\!=\!\!&
 -\,2f_K^2 M_K^2\,(\,1-\delta_K\,) \,,
\end{eqnarray}
where $\hat m\equiv(m_u+m_d)/2$ is the isospin average of the light up- and
down-quark masses, and $\delta_\pi$ as well as $\delta_K$ comprise the
higher-order chiral corrections. The left-hand sides of eqs.~\eqn{gmor1} and
\eqn{gmor2} just correspond to minus $\Psi(0)$ in the pseudoscalar case with
$(ud)$ and $(sq)$ quark flavours respectively. At the next-to-leading order,
these corrections have been found to be \cite{gl:84}:
\begin{equation}
\label{delpiK}
\delta_\pi = 4\,\frac{M_\pi^2}{f_\pi^2}\,(2L_8^r-H_2^r)
\quad \mbox{and} \quad
\delta_K = \frac{M_K^2}{M_\pi^2}\;\delta_\pi \,.
\end{equation}

The results for $H_2^r$ and $L_8^r$, obtained in the last section, now allow
for an estimation of the chiral corrections $\delta_\pi$ and $\delta_K$, which
take the values
\begin{equation}
\label{delpiKnum}
\delta_\pi = 0.047 \pm 0.017
\quad \mbox{and} \quad
\delta_K = 0.61 \pm 0.22 \,.
\end{equation}
With roughly 5\%, the size of the correction $\delta_\pi$ appears very
reasonable, whereas $\delta_K$ in the strange quark sector turns out rather
large, due to the enhancement factor $M_K^2/M_\pi^2$, although the
uncertainties are also quite big. If, on the other hand, the relation
$H_2^r=2L_8^r$ which follows from scalar resonance saturation \cite{egpr:89}
would be correct, both chiral corrections $\delta_\pi$ and $\delta_K$ would
have to vanish.

The knowledge of the light quark masses also allows for a determination of
the light quark condensate from the first GMOR relation \eqn{gmor1}. Using the
recent result $(m_u+m_d)(2\,\gev)=8.1\pm 1.4\;\mev$ \cite{jop:01}, together
with the estimate \eqn{delpiKnum} for $\delta_\pi$, one arrives at
\begin{equation}
\label{qq}
\langle\Omega|\bar qq|\Omega\rangle(2\,\gev) = -\,(267\pm 16\;\mev)^3 \,,
\end{equation}
which can be considered as an update of previous determinations of the
light-quark condensate $\langle\bar qq\rangle$. Since it is still more common
to quote the quark condensate at a scale of $1\;\gev$ the corresponding value
$\langle\bar qq\rangle(1\,\gev)=-\,(242\pm 15\;\mev)^3$ is also provided.
The latter results are in good agreement with direct determinations of
$\langle\bar qq\rangle$ from QCD sum rules \cite{dn:98} and lattice QCD
\cite{qqlat}.

\section{Conclusions}

From the structure of the low-energy theorem of eq.~\eqn{lot} for the scalar
(pseudoscalar) two-point function, it is possible to infer that the quark
condensate in $\chi$PT should be identified with the non-normal-ordered
vacuum average of quark-antiquark fields. Contrary to the corresponding
case of normal-ordered condensates, the product of quark mass times
non-normal-ordered condensate, and thus $\Psi(0)$, are not renormalisation
invariant. In $\chi$PT the dependence of $\Psi(0)$ on the short-distance
renormalisation scale and scheme is reflected in the dependence on the
unphysical low-energy constant $H_2^r$.

Because the low-energy constant $H_2^r$ depends on the renormalisation at
short-distances, its value cannot be deduced from relations to physical
quantities alone. One possible quantity to determine $H_2^r$ is the SU(3)
flavour-breaking ratio of the quark condensates $\langle\bar ss\rangle/
\langle\bar qq\rangle$. The estimate presented in eq.~\eqn{H2r} has been
obtained on the basis of $\langle\bar ss\rangle/\langle\bar qq\rangle$ from
QCD sum rules for the heavy $B$- and $B_s$-mesons, making additional use of
results on the ratio $f_{B_s}/f_B$ from lattice QCD.

The obtained value for $H_2^r$ also allows to estimate the chiral corrections
to the GMOR relations of eqs.~\eqn{gmor1} and \eqn{gmor2}. It is found that
the first GMOR relation in the light up- and down-quark sector receives a
moderate negative correction of the order of 5\%, whereas the corresponding
correction in the strange-quark sector is enhanced by a factor $M_K^2/M_\pi^2$,
and thus turns out to be rather large. Together with values for the light quark
masses, the first GMOR relation makes it possible to determine the light quark
condensate $\langle\bar qq\rangle$, and an updated value has been presented in
the last section.

Since the low-energy constants of $\chi$PT contain the physics of energies
higher than the pseudoscalar meson masses, it is possible to derive effective
Lagrangians including higher resonance fields, which then yield contributions
to the chiral constants \cite{egpr:89}. For the physical constants $L_i^r$,
very reasonable estimates were obtained in this manner. For the unphysical
constant $H_2^r$, however, it was shown above that the result $H_2^r=2L_8^r$
from the resonance Lagrangian is not satisfied. It should already be clear
that this relation has to receive additional contributions, since $H_2^r$
depends on the short-distance renormalisation, whereas this is not the case
for $L_8^r$.

Nevertheless, the uncertainties in the ratio $\langle\bar ss\rangle/\langle
\bar qq\rangle$ are still sizeable, so that no definite conclusions can be
reached at this stage. Further work should be devoted to the extraction of
$\langle\bar ss\rangle/\langle\bar qq\rangle$ from QCD sum rules, particularly
paying attention to the proper definition of the quark condensate as well as
to the corresponding higher-order strange-mass corrections. In addition, an
analysis analogous to the one presented here at the next-next-to-leading
${\cal O}(p^6)$ in $\chi$PT could provide further useful information on the
questions raised above, since also the uncertainty resulting from these higher
orders plays an important role.

\begin{acknowledgments}
It is a great pleasure to thank J.~Prades, H.~G.~Dosch, A.~Pich and
E.~de~Rafael for helpful discussions. The author would also like to
thank the Deutsche Forschungsgemeinschaft for their support.
\end{acknowledgments}

\end{document}